\documentclass[journal]{IEEEtran}
\usepackage{cite}

\ifCLASSINFOpdf
   \usepackage[pdftex]{graphicx}
\else
\fi
\usepackage{amsmath}
\usepackage{array}

\ifCLASSOPTIONcompsoc
 \usepackage[caption=false,font=normalsize,labelfont=sf,textfont=sf]{subfig}
\else
 \usepackage[caption=false,font=footnotesize]{subfig}
\fi
\usepackage{url}

\usepackage{xcolor}

\begin{document}
%
\title{Longitudinal Pooling \& Consistency Regularization to Model Disease Progression from MRIs}
%
%
%

\author{\IEEEauthorblockN{Jiahong Ouyang,
Qingyu Zhao,
Edith V. Sullivan, 
Adolf Pfefferbaum,
Susan F. Tapert,\\
Ehsan Adeli, \IEEEmembership{Member,~IEEE,} and
Kilian M. Pohl}

\thanks{Manuscript received June 22, 2020; revised Sept 3, 2020, and Oct 21, 2020.} 
\thanks{Funding for this study was received from the U.S. National Institutes Health (NIH) grants AA010723, AA026762, AA021681, AA021690, AA021691, AA021692, AA021695, AA021696, AA021697, and MH113406. This study also benefited from Stanford Institute for Human-centered Artificial Intelligence (HAI) AWS Cloud Credit.}
\thanks{J. Ouyang is with the Department of Electrical Engineering, Stanford University, Stanford, CA 94305.}
\thanks{Q. Zhao, E.V. Sullivan, A. Pfefferbaum, E. Adeli, and K.M. Pohl are with the Department of Psychiatry and Behavioral Sciences, Stanford School of Medicine, Stanford, CA 94305.}
\thanks{S.F. Tapert is with the Department of Psychiatry, University of California San Diego, La Jolla, CA 92093.}
\thanks{E. Adeli is also with the Department of Computer Science, Stanford University, Stanford, CA 94305.}
\thanks{A. Pfefferbaum and K.M. Pohl are also with the Center for Biomedical Sciences, SRI International, Menlo Park CA, 94025.}
\thanks{Corresponding author: K.M. Pohl (email: kilian.pohl@stanford.edu).}
}

\markboth{Longitudinal Pooling \& Consistency Regularization to Model Disease Progression from MRIs}%
{Shell \MakeLowercase{\textit{et al.}}: Bare Demo of IEEEtran.cls for IEEE Journals}
%



\maketitle

\begin{abstract}
Many neurological diseases are characterized by gradual deterioration of brain structure and function. Large longitudinal MRI datasets have revealed such deterioration, in part, by applying machine and deep learning to predict diagnosis. A popular approach is to apply Convolutional Neural Networks (CNN) to extract informative features from each visit of the longitudinal MRI and then use those features to classify each visit via Recurrent Neural Networks (RNNs). 
Such modeling neglects the progressive nature of the disease, which may result in clinically implausible classifications across visits. To avoid this issue, we propose to combine features across visits by coupling feature extraction with a novel longitudinal pooling layer and enforce consistency of the classification across visits in line with disease progression. We evaluate the proposed method on the longitudinal structural MRIs from three neuroimaging datasets: Alzheimer's Disease Neuroimaging Initiative (ADNI, $N=404$), a dataset composed of 274 normal controls and 329 patients with Alcohol Use Disorder (AUD), and 255 youths from the National Consortium on Alcohol and NeuroDevelopment in Adolescence (NCANDA). In all three experiments our method is superior to other widely used approaches for longitudinal classification thus making a unique contribution towards more accurate tracking of the impact of conditions on the brain. The code is available at \url{https://github.com/ouyangjiahong/longitudinal-pooling}.
\end{abstract}


%
\IEEEpeerreviewmaketitle

\section{Introduction}
Longitudinal neuroimaging studies enable scientists to track the gradual effect of neurological diseases and environmental influences on the brain over time \cite{whitwell2008longitudinal}. Compared to cross-sectional studies, modeling temporal dependency captured by longitudinal MRIs is crucial to accurately quantify the aging trajectories of anatomical and functional brain organization \cite{aubert2013new,herting2018development,fraser2015systematic} and monitoring the progression of neurological disorders in individuals \cite{bernal2013statistical,bilgel2016multivariate,cui2019rnn}.


Analysis of longitudinal MRIs is traditionally based on statistical approaches testing hypotheses \cite{skup2010longitudinal}. In this scenario, each image is first reduced to a set of brain measurements (e.g., volumes of regions of interest (ROIs)) \cite{Schwarz2016}. 
The effect of a condition (such as Alzheimer's Disease (AD)) on the longitudinal trajectory of each brain measurement is then tested by statistical approaches \cite{st1989analysis}. For example, general linear models (GLM) are often used to compute the average developmental trajectories for each cohort of interest (e.g., normal controls or AD patients) \cite{fjell2009one,sabuncu2011dynamics,frings2012quantifying}. The heterogeniety within each cohort can be captured by computing subject-specific trajectories via linear mixed effect (LME) models  \cite{bernal2013statistical,bernal2013spatiotemporal}. Transitional methods, such as Markov models \cite{salazar2007shared,dwyer2014improved}, can then further refine the modeling of temporal dependency underlying these trajectories. However, these univariate statistical approaches underestimate group differences as they cannot model the multivariate relations within the high-dimensional MRI \cite{Habeck2010cell}. Even for multi-variate models, analyses typically rely on {\it a priori} selection of MRI measurements \cite{Westman2012} that sub-optimally reflect the complex brain structure and function encoded by MRIs. Furthermore, inference of the findings is also limited to a cohort as they do not apply to individuals \cite{Rosenberg2018}. 

With recent advances in machine learning, inferences on an individual level are accomplished by training classifiers to distinguish longitudinal MRIs of normal controls from those of a cohort of interest. These classifiers can predict group assignment for each subject based on measurements of all brain regions thus avoiding the need of selecting measurements {\it a priori}. The most commonly used approach is to apply support vector machine (SVM)  \cite{gray2012multi,zhang2012predicting,zhang2017alzheimer} to features encoding information of longitudinal trajectories. For example, Gray and colleagues \cite{gray2012multi} concatenated ROI features from different time points with the assumption that the number of time points is the same across all subjects. Dropping this assumption, automatic longitudinal feature selection approaches \cite{zhang2012predicting} first remove unrelated ROI features by cross-sectional analysis and then fit a linear regression model to estimate the changing rate of the selected features over time. Beyond using hand-crafted features from pre-defined ROIs, Zhang et al. \cite{zhang2017alzheimer} proposed a landmark-based  feature extraction method that directly estimates data-driven features from images. Other feature selection methods are based on sparse learning (such as a multi-task sparse representation classifier \cite{liu2013multi} and a sparse Bayesian LME learner \cite{sabuncu2015sparse}), which can select a sparse subset of measurements that significantly contribute to the model prediction. The subset of measurements is then viewed as biomarkers associated with the disease \cite{adeli2018chained}.

State-of-the-art learning models are based on deep learning. Instead of using hand-crafted measurements, these models often reduce each MRI of the longitudinal sequence to informative features via Convolutional Neural Networks (CNN) and use the features to predict cohort assignment at each visit via Recurrent Neural Networks (RNN) \cite{lipton2015learning,santeramo2018longitudinal,gao2019distanced,cui2019rnn,ghazi2019training}. This architecture has advanced the analysis of longitudinal medical images, such as diagnosis \cite{cui2019rnn}, tissue segmentation \cite{bai2018recurrent}, and predicting disease progression \cite{lee2019predicting}. For example, Gao et al. \cite{gao2018brain} extracted features separately from each slice of a MRI via 2D CNN and then grouped features by Bag-of-Words before feeding them into the RNN layer. The RNNs themselves commonly consist of Long Short-Term Memory (LSTM) \cite{hochreiter1997long} and Gated Recurrent Units (GRU) \cite{cho2014learning}. For example, Cui et al. \cite{cui2019rnn} captured the temporal dependencies through a bidirectional GRU correlating information at each visit with pre- and proceeding visits. Others explicitly modeled the time interval between visits by injecting that information into the LSTM \cite{gao2019distanced,santeramo2018longitudinal}. After each RNN classification, the corresponding visit-specific assignments can be reduced to a single label by, for example, simply relying on the assignment of the last visit  \cite{santeramo2018longitudinal,gao2019distanced,cui2019rnn}, averaging across all visits \cite{xu2019deep}, or adding a linear layer on top of the RNN \cite{che2018recurrent} that weighs the importance of each visit for the final decision making \cite{wang2019toward}. 

Shortcomings of these deep learning approaches arise from their implementations of RNN, which, for example, analyze the longitudinal MRI in one temporal direction by performing inference at a visit only considering that and prior visits \cite{aghili2018predictive,pham2017predicting}. Such unidirectional assumption is suitable for applications related to real-time computer-aided diagnosis, which aim to assess the current visit without observing the future. In many neuroscience applications, however, the goal is to quantify group differences by accounting for all visits in the longitudinal sequence such as in longitudinally consistent segmentation \cite{Reuter2012}, registration \cite{Wu2011registration}, and network estimation \cite{zhao2019LICA}. While bidirectional RNNs could potentially alleviate this problem, they suffer from the `vanishing gradient' problem \cite{pascanu2013difficulty}. Another critical issue overlooked by current implementations of RNN models is that they do not explicitly model progression of a condition, such as neurocognitive decline. The RNN classifications across visits can thus be clinically implausible such as predicting a transition from the irreversible AD to Mild Cognitive Impairment (MCI). To produce longitudinally and clinically consistent labelling, we propose to add two novel components to the CNN and RNN framework: a {\it longitudinal pooling layer} before the RNN  to account for the features of all visits and a {\it consistency loss function} to regularize the RNN classifications through explicitly modelling irreversible conditions, such as in the case of AD. Both these two components can be generalized and plugged into models including but not limited to the combination of CNN and RNN. Specifically, longitudinal pooling layer can be adopted when the model needs to exploit the temporal dependency in features from multiple time points, while the consistency loss can be utilized when the trend in the longitudinal predictions are needed. 

Common approaches to account for the information across visits are feature concatenation \cite{nie20163d} and fully connected networks \cite{kamnitsas2015multi}. However, these models require the number of visits to be the same for each longitudinal scan, which is generally not the case for neuroscience studies \cite{matta2018making}. A potential approach for modelling varying number of visits across subjects is social pooling \cite{alahi2016social}, which was originally created to predict the spatial trajectories of pedestrian. Inspired by this model, our design of the longitudinal pooling layer generates (for each visit) a compact representation of MRI features of the current and proceeding visits before feeding them into the RNN. The classifications produced by the RNN are then regularized by a consistency loss function, which strongly discourages changes in cohort assignment of subjects with irreversible conditions, such as AD. For example, the function penalizes a decrease in confidence in labelling AD at a visit if the subject was labelled as such in prior ones. By doing so,  the sequence of scalar confidence values (across visits) has the potential to refine categorical modelling of disease progression, such as healthy, MCI, and AD. 

We evaluate our method on three longitudinal T1-weighted MRI datasets: 404 subjects from ADNI to analyze the progression of Alzheimer's disease (AD), 603 subjects to distinguish healthy subjects from those diagnosed with Alcohol Use Disorder (AUD), and 255 no-to-low drinking youths (ages 14 to 16 at baseline) of the National Consortium on Alcohol and NeuroDevelopment in Adolescence (NCANDA) to identify the ones that transition to heavy drinkers during early adulthood (i.e., age 18 years or older). On these data sets,  the accuracy of our proposed architecture is higher than alternative models without the longitudinal pooling and consistency layers. Finally, we derive voxel-wise saliency maps for the learned models to identify critical brain regions contributing to the classification. These regions converge with current understanding in neuroscience literature regarding the specific brain disorders or conditions under investigation. 

\section{Method}
Let $\{\textbf{x}_t^{(s)} | t=1,...,m_s\}$ be the sequence of structural MRIs $\textbf{x}_t^{(s)}$ of subject $s$ acquired over $m_s$ visits. Our deep learning model learns to predict binary labels $\{y_t^{(s)} | t=1,...,m_s\}$, which classify a subject $s$ at each visit $t$ being a control or belonging to the cohort of interest (e.g., AD). The total number of visits $m_s$ may vary across subjects. We make the simplifying assumption that the interval between visits is the same and that the same acquisition protocol is used throughout the study. We denote the entire dataset with $\mathcal{S}$.

\begin{figure*}[t]
\centering
\includegraphics[width=0.85\textwidth]{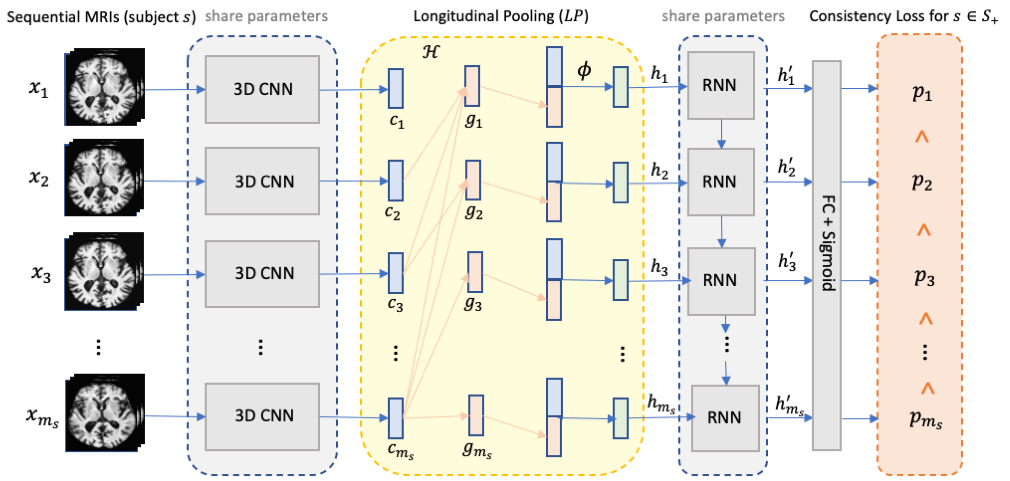}
\caption{The overview of the proposed method with the CNN+RNN backbone highlighted in gray. The CNN and RNN components share the same set of network parameters across time points. We introduce two innovative layers to the backbone: the longitudinal pooling layer (yellow component) augments the features derived by the 3D CNN at a visit with those of future visits and the consistency loss (orange) encourages higher prediction confidence  as time progresses for the subjects from the cohort of interest $\mathcal{S}_+$  (e.g., AD).}
\label{fig:overview}
\end{figure*}

\begin{figure}[!t]
\centering
\includegraphics[width=0.65\linewidth]{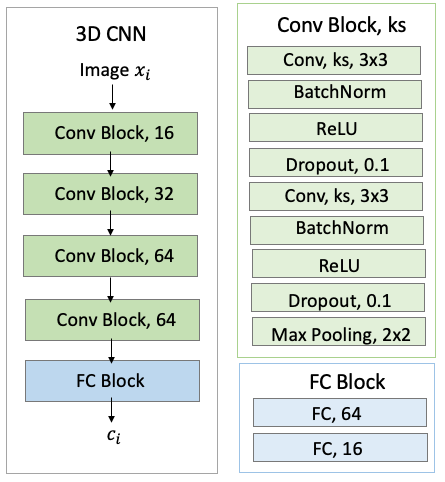}
\caption{Detailed structures of 3D-CNN. Green blocks denote convolutional (Conv) layers. Blue blocks denote fully connected (FC) layers. `ks' denotes kernel size.} 
\label{fig:cnn_arch}
\end{figure}

To predict the label $y_t^{(s)}$ from $\textbf{x}_t^{(s)}$, our method is based on a 3D-CNN coupled with a sequence-to-sequence RNN  (gray components in Fig. \ref{fig:overview}). Specifically, a 3D-CNN is independently applied to the MRI $\textbf{x}_t^{(s)}$ of each visit $t$ of the longitudinal sequence in order to extract informative features\footnote{Functions are typeset in italics.}
\begin{equation}
\textbf{c}_t^{(s)} := \text{\textit{CNN}}(\textbf{x}_t^{(s)}).    
\end{equation}
The specific design of our 3D-CNN is shown in Fig. \ref{fig:cnn_arch}. Note, the same 3D-CNN (i.e., set of weights) is used across visits in order for the model to accurately track the trajectories of features. This tracking is done by a RNN based on Gate Recurrent Unit (GRU) \cite{chung2014empirical}. GRU consists of an update gate and a reset gate that respectively determine how much of the past information needs to be passed on or forgotten. Let $\textbf{h}_t^{(s)}$ be the input, $\textbf{z}_t^{(s)}$ be the update gate vector, $\textbf{r}_t^{(s)}$ the reset gate vector, and $\textbf{h}_t^{\prime(s)}$ be the output of the RNN unit, then the network $\textbf{h}_t^{\prime(s)} := \text{\textit{RNN}}(\textbf{h}_t^{(s)}; \textbf{W}_r)$ is defined as 
\begin{align}
\textbf{z}_t^{(s)} :=& sigmoid(\textbf{W}_{rz} \textbf{h}_t^{(s)} + \textbf{U}_{rz} \textbf{h}_{t-1}^{\prime(s)})\nonumber\\
\textbf{r}_t^{(s)} :=& sigmoid(\textbf{W}_{rr} \textbf{h}_t^{(s)} + \textbf{U}_{rr} \textbf{h}_{t-1}^{\prime(s)})\\
\textbf{h}_t^{\prime(s)} :=&\textbf{z}_t^{(s)} \odot \textbf{h}_{t-1}^{\prime(s)} +\nonumber \\
& (1 - \textbf{z}_t^{(s)}) \odot tanh(\textbf{W}_{rh} \textbf{h}_t^{(s)} +\textbf{U}_{rh}(\textbf{r}_t^{(s)} \odot \textbf{h}_{t-1}^{\prime(s)})) \nonumber
\end{align}
where $\textbf{W}_{rz}, \textbf{W}_{rr}, \textbf{W}_{rh}, \textbf{U}_{rz}, \textbf{U}_{rr}, \textbf{U}_{rh}$ form the parameters $\textbf{W}_r$ of the GRU cell. To turn the output of each RNN cell into a binary label, $\textbf{h}_t^{\prime(s)}$  is fed into a fully connected (FC) layer followed by a sigmoid activation layer. We propose to extend this framework by adding a {\it Longitudinal Pooling} layer between the CNN and RNN, and feeding the binary label into a  {\it Consistency Loss function}. The remainder of this section describes these layers and the training of the method in detail. 

\subsection{Longitudinal Pooling (LP)}  
Inspired by social pooling \cite{alahi2016social}, the LP layer augments the features $\textbf{c}_t^{(s)}$ derived by the 3D CNN for subject $s$ at visit $t$ with those of future visits via a pooling operation $\mathcal{H}(\cdot)$, i.e.,  
\begin{align} 
    \textbf{g}_t^{(s)}:= 
   \begin{cases}
    \mathcal{H} (\{ \textbf{c}_u^{(s)} | u > t \}) & t < m_s\\
    \textbf{c}_t^{(s)}, & t = m_s
    \end{cases}~. 
    \label{eqn:pooling}
\end{align}
For simplicity, $\mathcal{H}(\cdot)$ computes the average. The last visit does not have proceeding visits to pool from, so we use $\textbf{c}_{m_s}^{(s)}$ itself. 

For each visit, the features $\textbf{c}_t^{(s)}$ of the current visit are concatenated with the pooling embedding $\textbf{g}_t^{(s)}$, which are then applied to a fully connected layer $\phi(\cdot)$ (i.e, a linear function with \texttt{tanh} activation and weights $\textbf{W}_f$) to determine the augmented hidden state
$\textbf{h}_t^{(s)} :=  \phi([\textbf{c}_t^{(s)}, \textbf{g}_t^{(s)}]; \textbf{W}_f)$. Lastly, $\textbf{h}_t^{(s)}$ becomes the new input to the RNN layer. Note, LP can be easily generalized to the scenario of multiple RNN layers. 
In this case, the pooling operation defined with respect to Eq. \eqref{eqn:pooling} and $\textbf{h}_{t}^{(s)}$ is performed on $\left\{\textbf{h}_{t}^{\prime(s)},\ldots, \textbf{h}_{ms}^{\prime(s)}\right\}$  instead of $\left\{\textbf{c}_t^{(s)},\ldots, \textbf{c}_{ms}^{(s)}\right\}$.

\subsection{Consistency Loss} 
To explicitly model irreversible conditions, such as in the case of AD, we encourage our approach to predict with higher confidence $p_t^{(s)}$ for subjects assigned to the cohort of interest $\mathcal{S}_+$  (e.g., AD) as time progresses. In other words, the consistency loss function expects $p_1^{(s)} \leq p_2^{(s)}...\leq p_m^{(s)}$ for $s \in \mathcal{S}_+$ and penalizes classifications violating this rule: 
\begin{equation}
    L_{cons} := \sum_{s \in \mathcal{S}_+} \sum_{1 \leq i  < m_s} \lfloor p_i^{(s)} - p_{i+1}^{(s)} \rfloor_{+}~, 
    \label{eqn:consistency}
\end{equation}
where $\lfloor \cdot \rfloor_{+}$ sets negative values to 0. Note, this consistency loss is a soft restriction, which only encourages but does not strictly enforce the monotonicity of $p_i^{(s)}$. This practice is more computationally stable than enforcing the hard inequality constraint in general deep learning frameworks \cite{marquez2017imposing}. Lastly, the consistency loss ignores subjects not belonging to $\mathcal{S}_+$ (e.g., health controls) as the condition does not apply to them.

\subsection{Objective Function and Training Strategy}
The final training stage aims to minimize an objective function consisting of the consistency loss $L_{cons}$ of Eq. \eqref{eqn:consistency}, a weighted binary cross entropy, and a term regularizing the weights $\textbf{W}=\{\textbf{W}_r, \textbf{W}_f\}$ of all recurrent and linear layers. Let  $\parallel \cdot \parallel_2$ be the L$_2$-norm and the binary entropy be 
\begin{equation}
E:= \sum_{s \in \mathcal{S}} \sum_{i=1}^{m_s} \big(w_{pos} \cdot y_i^{(s)} \log(p_i^{(s)}) + (1 - y_i^{(s)}) \log(1- p_i^{(s)})\big)
\end{equation} 
with the parameters $w_{pos}$ balancing the influence of the cohort of interest  $S_+$ over the controls ($S \backslash S_+$). Then the objective function is defined as
\begin{equation}
    L  := -E + \lambda_{cons} L_{cons} + \lambda_{reg} \parallel \textbf{W} \parallel_2^2 ~,
\end{equation}
where $\lambda_{cons}$ and $\lambda_{reg}$ weigh the importance of the consistency loss and regularization loss over the weighted binary entropy. 

As jointly training CNN and RNN from scratch can easily lead to overfitting due to the large number of parameters, we first pre-train the convolutional blocks (green blocks in Fig. \ref{fig:cnn_arch}) of the 3D-CNN as proposed in \cite{cui2019rnn}. We do so by training a cross-sectional CNN-based classifier (i.e. applying a linear classifier to $\textbf{c}_t^{(s)}$) on all available training images while discarding their longitudinal dependencies. Based on the pre-trained network parameters in the convolutional blocks, we re-initialize the parameters in the FC blocks and jointly train our CNN+RNN model with the recurrent and longitudinal pooling layers.
Alternative approaches for pre-training the CNN (not explored here) are to apply unsupervised representation learning frameworks (such as auto-encoders \cite{suk2015latent}) or self-supervised learning frameworks, such as patch-based \cite{doersch2015unsupervised,noroozi2017representation} and distortion-based approaches \cite{gidaris2018unsupervised}.

\section{Experiments}
We applied this method to three longitudinal neuroimaging datasets: a subset of ADNI consisting of 214 normal controls and 190 AD patients, a dataset acquired by us (AP and EVS) of 274 normal controls and 329 AUD patients, and 255 no-to-low drinking youths (ages 14 to 16 at baseline) of NCANDA, of which 69 transitioned to heavy drinkers once they became adults. For each dataset, we compared the classification accuracy of the proposed model with several other widely used baselines and visualized the saliency of our model to identify brain regions critically contributing to the classification. The remainder of this section describes the experimental setup and findings on the three datasets in detail.

\subsection{Experiment Setting}
\subsubsection{Data Preprocessing}
In line with our prior cross-sectional study \cite{zhao2019confounder,adeli2020deep}, all longitudinal MRIs in the following experiments were first preprocessed by a pipeline composed of denoising, bias field correction, skull striping, affine registration to a template, and re-scaling to a $64\times64\times64$ volume \cite{zhao2019confounder}. At the sacrifice of image resolution, the downsampling enables the design of a compact CNN model with a relatively small number of network parameters, which can effectively boost training speed and avoid the vanishing gradient problem in training RNN models \cite{pascanu2013difficulty}. Note, while theoretically not required for the training of CNN-based methods, 
these processing steps generally result in faster convergence, less overfitting, and more accurate classification \cite{wen2020convolutional}.

Next, we split the data set into  training, validation, and testing sets. After randomly selecting 10\% subjects as the validation set, the remaining subjects were split into 5 folds for cross-validation while keeping the proportion of $\mathcal{S}_+$ the same in  each fold. For each testing run, the corresponding training data were augmented by applying random 3D poses (rotation and shifting) to the  longitudinal MRIs, where the same pose was applied to each MRI of the longitudinal sequence. Furthermore, we flipped hemispheres based on the assumption that the studied condition effected the brain bilaterally, which is the case for most neurological diseases. This process increased the size of the training set by a factor of 10. Based on this augmented data set from the training folds, we performed the pre-training of CNN and the joint training of CNN+RNN.

\subsubsection{Model Architecture \& Hyperparameters}
As shown in Fig. \ref{fig:cnn_arch}, the 3D CNN contained 4 convolutional blocks connected by $2 \times 2 \times 2$ 3D MaxPooling. Each block consisted of two stacks of $3 \times 3 \times 3$ 3D convolution (16/32/64/64 as number of channels for the 4 blocks), Batch Normalization, ReLU, and dropout layers. The resulting 512-D features were connected to two fully connected layers (FC) with \texttt{tanh} activation producing a 16-D feature ($\textbf{c}_t$) as the input of RNN. To reduce the risk of overfitting, the RNN implementation was constructed by a GRU layer with 16 hidden units. In comparison to the commonly used LSTM \cite{hochreiter1997long}, GRU not only requires training of fewer network parameters but also trains faster \cite{chung2014empirical}. We set $\lambda_{reg}=0.02$ and $\lambda_{cons}=2.0$ as they resulted in the highest accuracy score on the validation set.  

\subsubsection{Evaluation Metrics \& Visualization Method}
While the 5 folds were split based on subjects (or longitudinal MRIs), we measured the accuracy of our proposed method ({\bf CNN+RNN+LP+CL}) with respect to each individual MRI of the longitudinal sequence in the testing fold. Specifically, we measured the sensitivity (SEN), specificity (SPE), Area Under the Curve (AUC) in percent, and the balanced accuracy (BACC) \cite{park2018sr}, where the classification accuracy for each cohort was balanced with respect to the total number of visits (or MRIs) across all subjects of the cohort. The standard deviation of BACC was computed across 5 folds for each method. To put those accuracy scores in perspective, we repeated the experiment for alternative classification models. First, we applied a  cross-sectional {\bf CNN} to each MRI of a longitudinal scan \cite{esmaeilzadeh2018end} (i.e., discarding the temporal relationship). The CNN baseline was also used in \cite{liu2014multimodal,hosseini2016alzheimer} for normal control vs. AD classification on the ADNI dataset. Next, we added an average pooling layer to the CNN baseline by concatenating the CNN features of a visit with the average of the CNN features across all the other visits ({\bf CNN+AP}). The basic longitudinal  approach  was the standard {\bf CNN+RNN} that was previously applied to the ADNI dataset in \cite{gao2018brain,cui2019rnn}.  On top of this baseline, we also added the LP layer ({\bf CNN+RNN+LP}). In addition, we also applied two state-of-the-art longitudinal models: {\bf CNN+biRNN} \cite{cui2019rnn} and {\bf CNN+tRNN} \cite{santeramo2018longitudinal}. biRNN captured information from all visits based on a  bidirectional GRU layer and tRNN modelled the potentially imbalanced time interval between successive visits within a  LSTM cell. To make the results comparable across methods, each approach used  the same CNN architecture and training strategy.

In addition to reporting accuracy scores, we visualized brain regions for driving the model decision by computing the saliency maps for each subject and visit. Extending the original visualization approach \cite{simonyan2013deep} to a longitudinal setting, we computed voxel-wise partial derivatives (measure of saliency values) of the classification at each visit with respect to each MRI in the longitudinal series. For example, a subject with 5 visits will have 5 $\times$ 5 saliency maps, where the map at the $i^{th}$ row and $j^{th}$ column shows the important brain regions of the MRI at the $j^{th}$ visit with respect to the classification at $i^{th}$ visit. Thus, the lower triangular of the 5 $\times$ 5 matrix represents how the classification at each visit depends on prior ones, the diagonal represents the current visit, and the upper triangular reveals the relevance of proceeding visits. The saliency values were re-normalized by 98\% of the maximum value within each subject to account for potential outliers in estimating partial-derivatives. The mean across the saliency maps of all subjects was overlaid with the SRI24 atlas \cite{rohlfing2010sri24} to relate brain areas most relevant to the specific brain disorder (or condition).

\subsection{Application to the ADNI Data Set}
\begin{table}[!t]
\centering
\begin{tabular}{lcccc}
Method &  BACC $\pm$ std & SEN & SPE & AUC\\
\hline
CNN &  86.1 $\pm$ 1.83 & 86.8 & 85.4 & 91.3\\
CNN+AP & 87.8 $\pm$ 2.55  & 88.2 & 87.4 & 91.3\\
\hline
CNN+RNN & 89.5 $\pm$ 1.41 & 88.7 & 90.3 & 91.4\\
CNN+RNN+LP & 90.0 $\pm$ 1.44 & \textbf{89.2} & 90.8 & \textbf{91.5}\\
CNN+RNN+LP+CL & \textbf{90.4} $\pm$ 1.51 & 88.9 & 91.9 & \textbf{91.5}\\
\hline
CNN+biRNN \cite{cui2019rnn} & 89.6 $\pm$ 1.48 & 86.8 & \textbf{92.4} & \textbf{91.5}\\ 
CNN+tRNN \cite{santeramo2018longitudinal} & 89.7 $\pm$ 1.67 & 88.8 & 90.6 & 91.4

\end{tabular} 
\vspace{10pt}
\newline
\centering
\begin{tabular}{cc}
\rotatebox{90}{\hspace{-8.5mm}\tiny{CNN+RNN+LP+CL}} \! &
\begin{tabular}{c|ccc}
Visits & NC/AD & BACC $\pm$ std \\
\hline
1+ visit & 188/169 & 89.8 $\pm$ 1.48\\
2+ visits & 163/103& 90.1 $\pm$ 1.53\\
3+ visits & 138/78& 90.6 $\pm$ 1.59\\
4+ visits & 109/52 & 90.6 $\pm$ 1.46\\
5+ visits & 73/0 & 91.8 $\pm$ 1.49
\end{tabular}
\end{tabular} \\
\vspace{10pt}

\caption{ADNI Dataset. Top: Comparison across methods on NC vs. AD classification; Bottom: Balanced Classification Accuracy of the proposed method dependent on the number of visits}\label{adni-a}
\end{table}

\begin{table}[!t]
\centering
\begin{tabular}{ccccc}
Method & Type & Modalities & NC/AD & BACC\\
\hline
\multicolumn{4}{l}{Longitudinal}\\
Zhang et al.\cite{zhang2017alzheimer} & N & MRI, PET & 207/154 & 88.3 \\
Gray et al.\cite{gray2012multi} & N & MRI, PET & 54/50 & 88.4 \\
Aghili et al.\cite{aghili2018predictive} & R & MRI, PET, CSF, genetic & 521/336 & 95.3 \\
Gao et al.\cite{gao2018brain} & C+R & MRI & 154/111 & 89.5 \\
Cui et al.\cite{cui2019rnn} & C+R & MRI & 229/198 & 91.3 \\
Proposed & C+R & MRI & 188/169 & 90.4 \\
\end{tabular} 
\caption{Comparison of the proposed method with other traditional methods and deep learning based methods on ADNI dataset. `C' denoted CNN, `R' denoted RNN, and `N' denoted non-deep-learning methods.}
\label{adni-compare}
\end{table}

Based on all successfully processed T1-weighted MRIs of the ADNI data set, we applied the proposed method to distinguish 214 normal controls (NC; age: 75.57 $\pm$ 5.06 years, 108 male / 106 female) from 190 patients diagnosed with Alzheimer's disease (AD; age: 75.17 $\pm$ 7.57 years, 104 male/86 female). There was no significant age difference between the NC and AD cohorts (p=0.55, two-sample \textit{t}-test).  The number of visits $m_s$ varied from 1 to 5 covering the first two years of the ADNI study with 6-month intervals.

According to the accuracy scores listed in Table \ref{adni-a} (top), each implementation recorded a sensitivity that was similar to its specificity indicating that the classification accuracy was fairly balanced across the two cohorts. However, the accuracy scores were quite different across implementations with CNN recording the lowest BACC of 86.1$\%$. The accuracy slightly improved when adding subject-level average pooling to the model (CNN+AP) but was still lower than that of the longitudinal models. This indicates that omitting the temporal information when performing classification on longitudinal MRIs might result in suboptimal results. Of those longitudinal models, CNN+RNN recorded the lowest accuracy and AUC, implementations including LP achieved the highest AUC (91.5\%), and the highest BACC of 90.4\% was recorded for the model with both LP and consistency loss (CL). This accuracy score was also higher than the two alternative longitudinal models CNN+tRNN (89.7$\%$) and CNN+biRNN (89.6$\%$). The score for CNN+biRNN was also confirmed by the original publication of biRNN \cite{cui2019rnn} (i.e, 89.4$\%$ when applied to a similar subset of the ADNI data), which suggests that the proposed longitudinal pooling might be more suitable than biRNN in extracting information from proceeding visits. Only our proposed model was significantly more accurate than the CNN baseline (p$<$0.005 according to DeLong's test \cite{delong1988comparing}).


To further show the impact of the consistency regularization, we added 137 ADNI subjects labelled as progressive MCI (pMCI) to the training data set; pMCIs transitioned from MCI to AD during the study, so they potentially exhibit more pronounced progression of cognitive decline than the patients diagnosed as AD since the beginning. Correctly capturing disease progression can guide the model to focus on cognitive decline, which in turn improves the NC/AD classification. Therefore, our novel consistency loss was defined on all three cohorts including the pMCI subjects while the classification loss was confined to NC and AD subjects. Incorporating the pMCI subjects increased the BACC of NC and AD classification to 90.8$\%$. This result suggests that including the pMCI subjects in the consistency loss component regularizes the model such that the prediction scores for each time point are related to the state of cognitive decline among subjects and reflect the progressive nature of the disease.


To put the above accuracy scores in perspective, we compared our results with those reported by prior studies using traditional machine learning and deep learning (Table \ref{adni-compare}). In general, deep-learning-based methods tend to have higher or comparable prediction accuracy than traditional machine learning methods, suggesting that the data-driven feature extraction is more sensitive in revealing group differences compared to hand-crafted measurements. Among the three longitudinal studies that use CNN+RNN, the number of training samples may become a factor impacting the classification accuracy as the BACC increases with the size of the training set. Despite the discrepancy in experimental setups (e.g., number of samples and imaging modalities) across these studies, our proposed method still achieved relatively good performance among all the reported scores.

In Table \ref{adni-a} (bottom), we also recorded testing accuracy on sub-groups confined to the number of visits greater or equal to a certain threshold. We observed that subjects with more visits tend to have better accuracy. CNN+RNN+LP+CL achieved the highest accuracy of 91.8$\%$ when basing classifications on the longitudinal MRIs of subjects with 5 visits. This indicates that our model was effectively capturing the changes within the longitudinal MRI. In addition, for subjects that have fewer number of visits, the proposed method still outperformed CNN+AP, suggesting the superiority of the proposed longitudinal pooling over average pooling.

\begin{figure}[!t]
\centering
\includegraphics[width=0.95\linewidth]{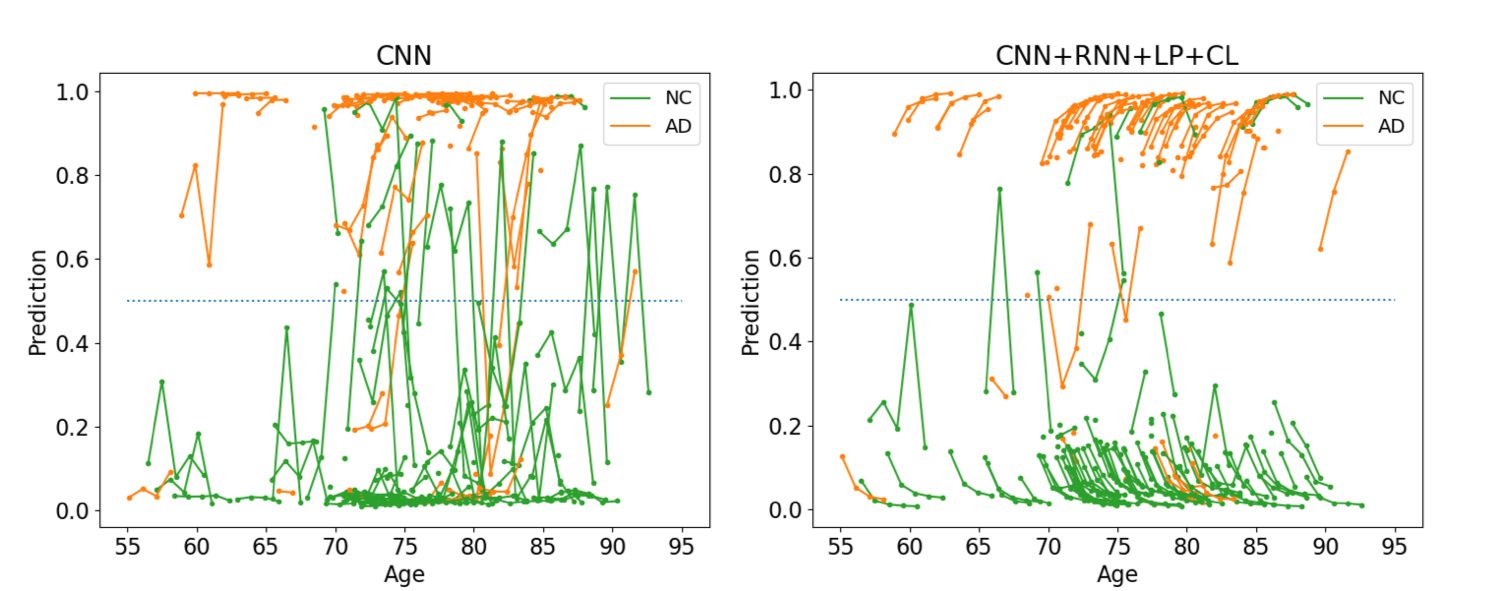}
\caption{Classifications by CNN and the proposed method on ADNI} 
\label{fig:vis-adni}
\end{figure}

Fig. \ref{fig:vis-adni} qualitatively confirms this finding. Many of the classifications generated by the cross-sectional CNN fluctuate between NC and AD as the lines (which connect classifications of the same subject) frequently cross 0.5 (dotted blue line). These clinically impossible transitions rarely happened for the classifications of our approach. The intra-subject classification scores for most AD patients increased with each visit suggesting that one might be able to use the scores for tracking disease progression. Although only applied to the AD patients, the consistency loss seemed to also regularize the prediction of NC subjects, which generally decreased with time and resulted in a clear separation between NC and AD cohorts.

\begin{figure}[!t]
\centering
\subfloat[Saliency matrix derived by regular CNN+RNN]{\includegraphics[width=0.8\linewidth]{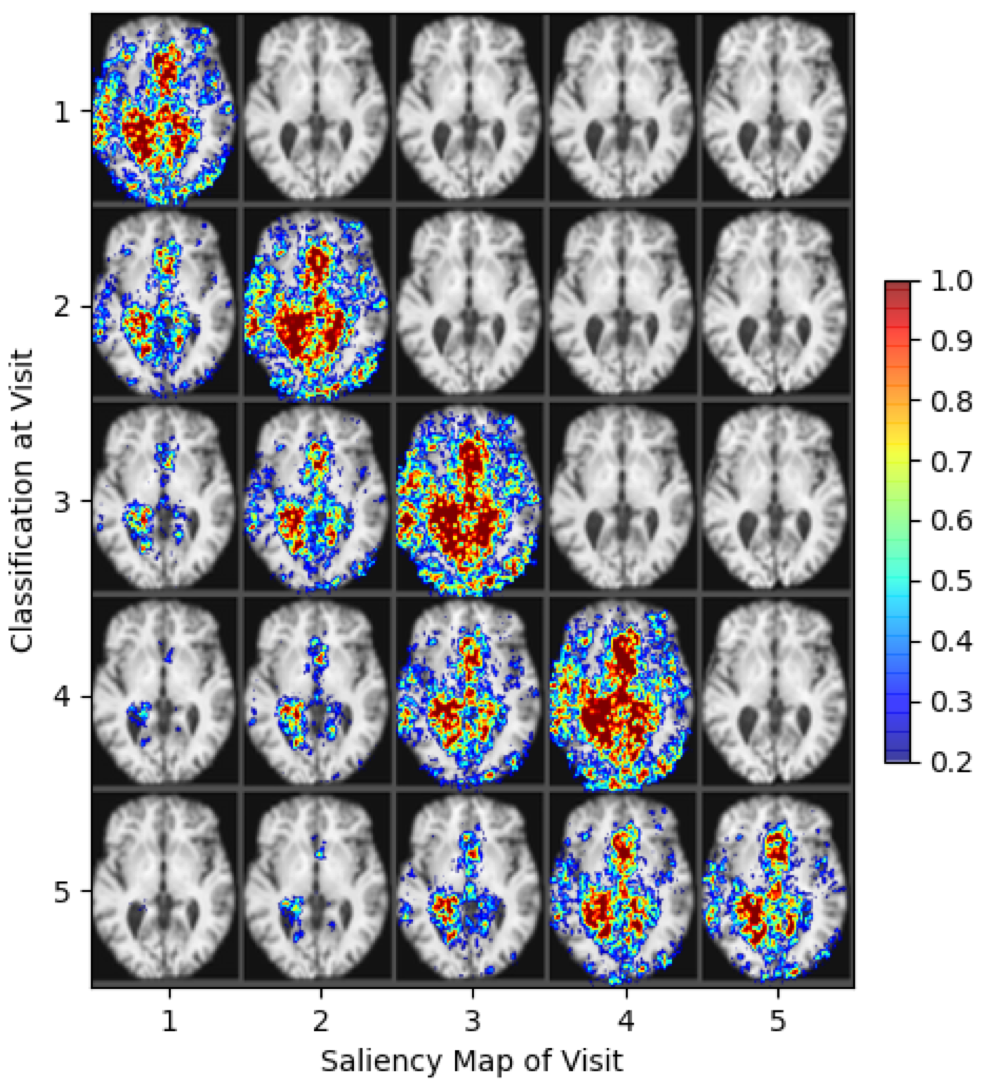}%
\label{fig:map-adni-matrix-rnn}}\\
\subfloat[Saliency matrix derived by the proposed model]{\includegraphics[width=0.8\linewidth]{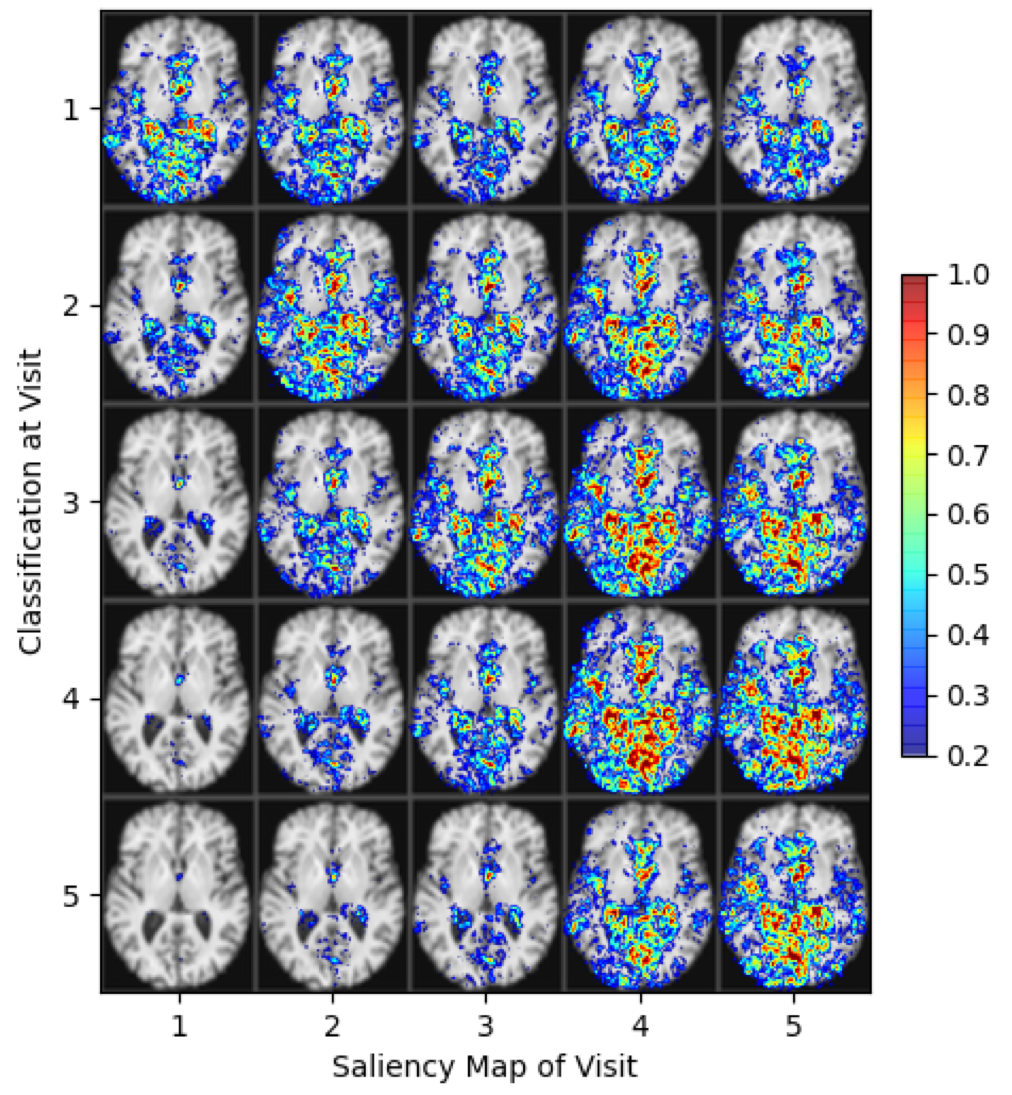}%
\label{fig:map-adni-matrix}}\\
\subfloat[Average saliency map of the proposed model]{\includegraphics[width=0.75\linewidth]{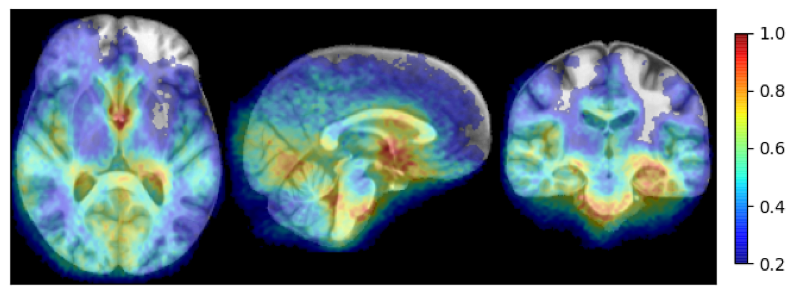}%
\label{fig:map-adni-mean}}
\caption{Saliency of controls versus AD}
\label{fig:map-adni}
\end{figure}

The salience maps of this classification model are shown in Fig. \ref{fig:map-adni}. The example of the salience matrix derived by CNN+RNN  (see Fig. \ref{fig:map-adni-matrix-rnn}) reveals that most salient regions are in the maps located at the diagonal, which indicates that the classification decision for each visit was mainly driven by the MRI of that visit alone. In comparison, the saliency matrix of the proposed model (Fig. \ref{fig:map-adni-matrix}) highlights salient regions across all visits slightly favoring proceeding visits for classifications at later visits (i.e., upper triangle). Lastly, Fig. \ref{fig:map-adni-mean} shows the average saliency map for the proposed model across all subjects. Areas of high saliency are the hippocampus and lateral ventricles, which are known to subserve memory consolidation and affected in early stages of AD \cite{alzheimer20172017}. High saliency was also found in the cerebral cortex potentially corresponding to brain atrophy and loss of brain mass\cite{Leung2013ad}.

\subsection{Identifying AUD based on Longitudinal MRIs}

AUD often causes gradual deterioration in the gray and white matter tissue \cite{Pfefferbaum14,Zahr2017}. In this experiment, we applied the proposed method to distinguish longitudinal T1-weighted MRIs (up to 4 visits) of 274 normal controls (NC; age: 47.3 $\pm$ 17.6, 136 male / 138 female) from those of 329 patients diagnosed with Alcohol Use Disorder (AUD; age: 49.3 $\pm$ 10.5, 100 male / 229 female). The study was approved by the institutional review boards of Stanford University School of Medicine and SRI International.

\begin{table}[!t]
\centering
\begin{tabular}{lcccc}
Method &  BACC $\pm$ std & SEN & SPE & AUC\\
\hline
CNN &  67.6 $\pm$ 3.54 & 62.9 & 72.2 & 70.9\\
CNN+AP &  68.7 $\pm$ 3.68 & 66.7 & 70.7 & 71.5\\
\hline
CNN+RNN & 69.0 $\pm$ 3.35 & 67.1 & 70.9 & 71.6\\
CNN+RNN+LP & \textbf{69.5} $\pm$ 3.32 & 67.6 & \textbf{71.4} & \textbf{71.8} \\
CNN+RNN+LP+CL & 69.3 $\pm$ 3.31 & \textbf{67.9} & 70.8 & \textbf{71.8}\\
\hline
CNN+biRNN \cite{cui2019rnn} & 68.9 $\pm$ 3.36 & 67.8 & 70.0 & 71.7\\ 
CNN+tRNN \cite{santeramo2018longitudinal} & 69.2 $\pm$ 3.42 & 68.6 & 69.8 & \textbf{71.8}
\end{tabular} 
\vspace{10pt}
\newline
\begin{tabular}{cc}
\rotatebox{90}{\hspace{-8.5mm}\tiny{CNN+RNN+LP+CL}} \! &
\begin{tabular}{c|ccc}
Visits & NC/AUD & BACC $\pm$ std \\
\hline
1+ visit & 244/293 & 65.0 $\pm$ 3.34\\
2+ visits & 124/150& 70.6 $\pm$ 3.32\\
3+ visits & 63/74& 77.1 $\pm$ 3.28\\
4+ visits & 30/29 & 79.5 $\pm$ 3.29\\
\end{tabular}
\end{tabular}
\vspace{10pt}
\caption{AUD Dataset. Top: Comparison across methods on NC vs. AUD classification; Bottom: Balanced Classification Accuracy of the proposed method dependent on the number of visits}\label{lab-a}
\end{table}

\begin{figure}[!t]
\centering
\includegraphics[width=0.95\linewidth]{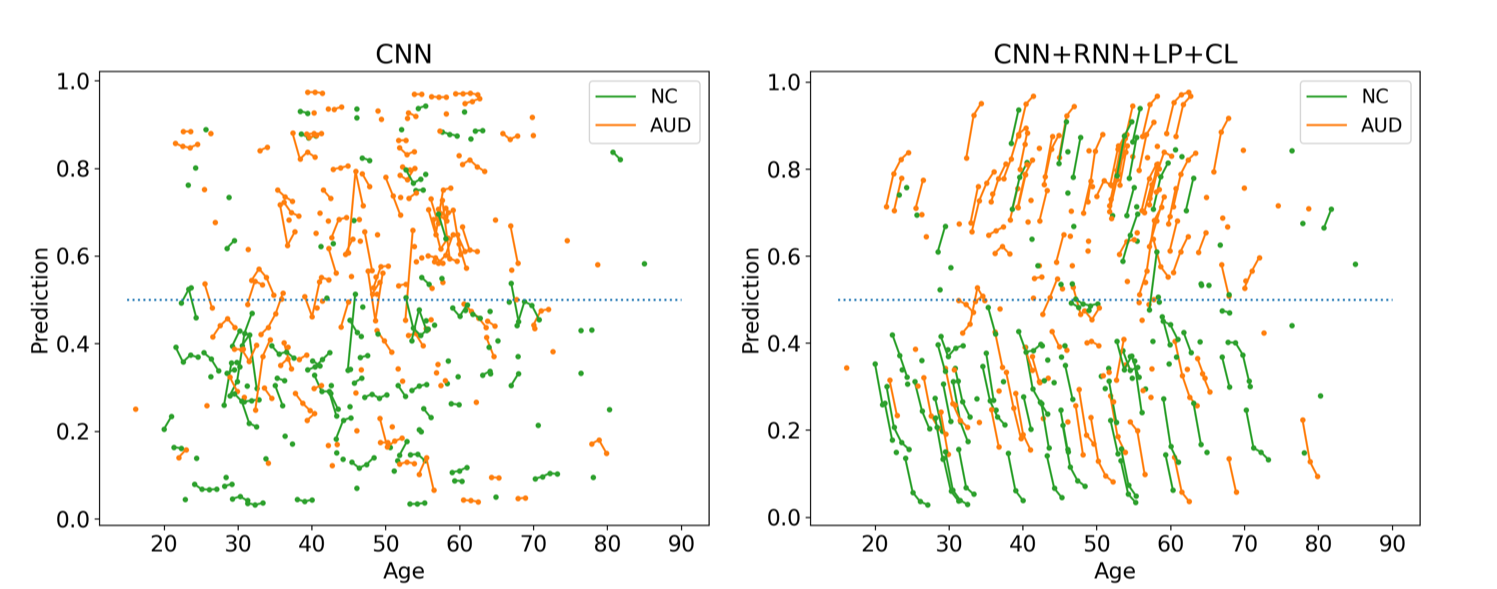}
\caption{Classifications by CNN and the proposed method on AUD} 
\label{fig:vis-lab}
\end{figure}

As it was the case in the ADNI experiment, all implementations (except CNN) recorded a sensitivity that was similar to its specificity and achieved accuracy scores significantly more accurate than chance (p$<$1e-5; Fisher Exact test \cite{fisher1922interpretation}); the longitudinal approaches were more accurate than the cross-sectional ones, and the accuracy of CNN+RNN+LP+CL improved with the number of available visits for classification (according to Table \ref{lab-a}). However, the accuracy scores were much lower than recorded for ADNI and the consistency loss did not improve the accuracy as CNN+RNN+LP and CNN+RNN+LP+CL had similar scores. One reason could be that even the cross-sectional CNN produced classifications that were relatively stable across visits (see Fig. \ref{fig:vis-lab}) so that the benefit of the consistency loss function was limited for this dataset. Another reason could be the relatively small number of subjects with more than two visits (Table \ref{lab-a}). Nevertheless, our proposed method was more accurate for subjects with more visits and achieved a BACC of 79.5\% for subjects with four visits. This result supported the advantage of longitudinal modeling proposed herein.


\begin{figure}[!t]
\centering
\subfloat[Saliency matrix derived by the proposed model]{\includegraphics[width=0.75\linewidth]{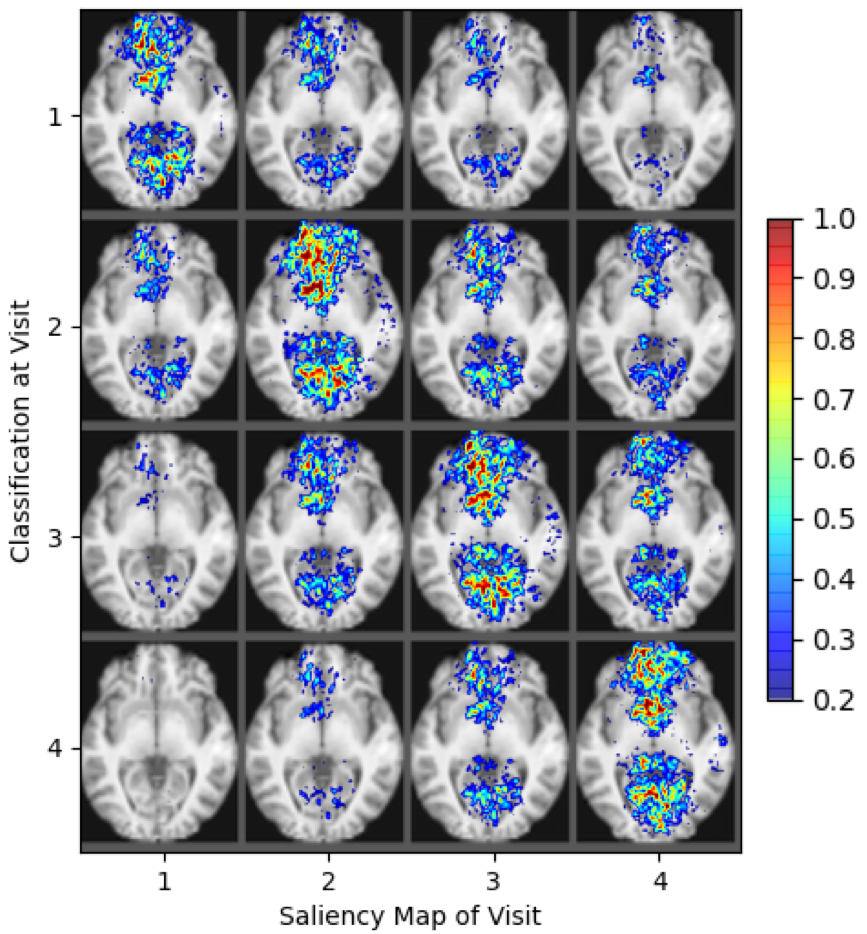}%
\label{fig:map-lab-matrix}}\\
\subfloat[Average saliency map of the proposed model]{\includegraphics[width=0.75\linewidth]{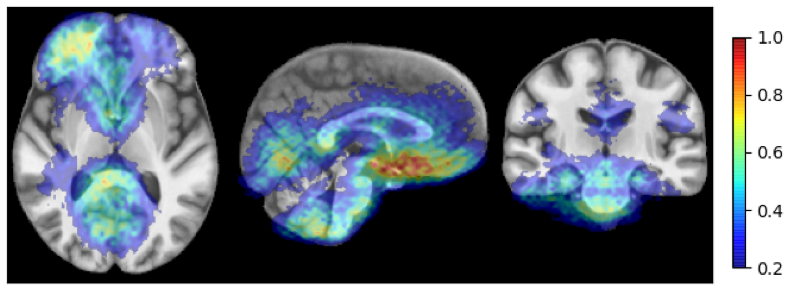}%
\label{fig:map-lab-mean}}
\caption{Saliency specific to AUD}
\label{fig:map-lab}
\end{figure}

Similar to the ADNI experiment, the example of a saliency matrix (Fig. \ref{fig:map-lab-matrix}) derived by our model recorded 
high saliency in maps located at the off-diagonal implying that the proposed model utilized information from both preceding and proceeding visits. Unlike the ADNI experiment, the average saliency map (Fig. \ref{fig:map-lab-mean}) now concentrates on specific brain regions frequently linked to AUD, such as the calcarine and lingual regions of the occipital lobe \cite{MO2014}, the orbitofrontal cortex \cite{Moorman2018}, and the cerebellum \cite{zhao2019cb}.

\subsection{Application to the NCANDA Data Set }

One aim of NCANDA is to study the brains of adolescents before drinking appreciable levels of alcohol in order to identify predictive phenotypes associated with heavy drinking \cite{brown2015national}. Based on the public data release NCANDA\_PUBLIC\_4Y\_STRUCTURAL\_V01 (DOI: 10.7303/syn22216457; distributed to the public according to the NCANDA Data Distribution agreement \url{https://www.niaaa.nih.gov/research/major-initiatives/national-} \url{consortium-alcohol-and-neurodevelopment-adolescence/ncanda-data )}, we applied our implementations to the longitudinal MRIs (up to 5 visits) of the 255 no-to-low drinking adolescents (124 boys / 131 girls) of this study that were 14-16 years old at baseline. During late adolescence to young adulthood (i.e, age 18 or older), 115 subjects remained no-low drinkers, 71 subjects transitioned to moderate drinkers, and 69 were heavy drinkers according to the adjusted Cahalan criteria \cite{pfefferbaum2018altered}. Our implementation now aimed to differentiate the heavy from no-to-low drinkers, while we reduced the risk of overfitting by adding the moderate drinkers to the heavy drinking cohort during training.

\begin{figure}[t]
\centering
\includegraphics[width=0.95\linewidth]{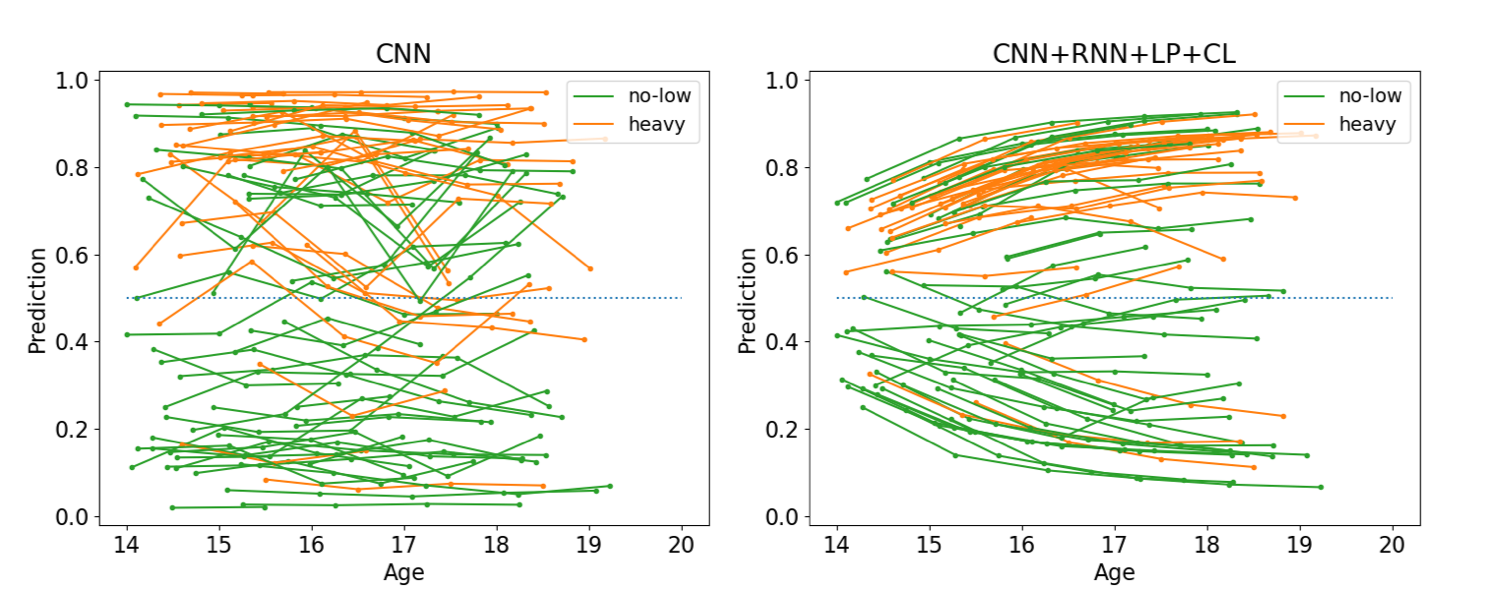}
\caption{Classifications by CNN and the proposed method on NCANDA} 
\label{fig:vis-ncanda}
\end{figure}

Again, findings of the previous experiments were largely confirmed. However, sensitivity and specificity were not as balanced as in previous experiments as all implementations favored the heavy drinking cohort. This resulted in the baseline CNN producing the highest AUC across all comparison methods. However, when using a non-informative operating point of 0.5 (threshold for class assignment), CNN+RNN+LP+CL achieved the highest accuracy with 72.7$\%$ and was significantly more accurate than the CNN baseline (p$<$0.001, DeLong's test). 
Similar to the ADNI experiments, the consistency loss layer improved the accuracy of the classifications. Moreover, the classification scores of the cross-sectional CNN implementation greatly varied across time, while this was not the case for the proposed implementation (see Fig. \ref{fig:vis-ncanda}). This underlines the importance of the pooling and consistency layer of our longitudinal CNN+RNN implementation, which produced classification scores that might be able to be used for tracking the impact of alcohol drinking on the adolescent brain. Unlike the previous two experiments, Table \ref{ncanda-a} shows that the classification accuracy was less correlated with the number of visits used for classification despite that the highest BACC was associated with subjects with 5 or more visits. This indicates that risk factors for heavy alcohol drinking potentially precede the neurodevelopment before age 14 years \cite{donovan2011childhood}. This finding is also echoed by the relatively ``flatter" classifications along age in Fig. \ref{fig:vis-ncanda} compared to previous plots for AD in Fig. \ref{fig:vis-adni} and AUD in Fig. \ref{fig:vis-lab}.

\begin{table}[t]
\centering
\begin{tabular}{lcccc}
Method &  BACC $\pm$ std & SEN & SPE & AUC\\
\hline
CNN &  68.8 $\pm$ 2.81 & 75.5 & 62.1 & \textbf{74.5}\\
CNN+AP &  71.0 $\pm$ 3.67 & 77.3 & 64.7 & 74.2\\
\hline
CNN+RNN & 70.8 $\pm$ 3.26 & 77.2 & 64.4 & 73.3\\
CNN+RNN+LP & 72.1 $\pm$ 3.23 & 80.4 & 63.8 & 73.4 \\
CNN+RNN+LP+CL & \textbf{72.7} $\pm$ 3.21 & \textbf{81.3} & 64.1 & 73.4\\
\hline
CNN+biRNN \cite{cui2019rnn} & 71.4 $\pm$ 3.25 & 78.3 & 64.5 & 74.1\\ 
CNN+tRNN \cite{santeramo2018longitudinal} & 71.8 $\pm$ 3.30 & 76.8 & \textbf{66.8} & 73.8
\end{tabular} 
\vspace{10pt}
\newline
\begin{tabular}{cc}
\rotatebox{90}{\hspace{-10.5mm}\tiny{CNN+RNN+LP+CL}} \! &
\begin{tabular}{c|ccc}
Visits & no-low/heavy & BACC $\pm$ std \\
\hline
1+ visit & 100/60 & 71.5 $\pm$ 3.23\\
2+ visit & 99/60& 72.7 $\pm$ 3.26\\
3+ visit & 93/56& 72.3 $\pm$ 3.25\\
4+ visit & 80/45 & 71.9 $\pm$ 3.18\\
5+ visit & 37/20 & 79.0 $\pm$ 3.18\\
\end{tabular}
\end{tabular}
\vspace{10pt}
\caption{NCANDA Dataset. Top: Comparison across methods on no-low vs. heavy drinking classification; Bottom: Balanced classification accuracy of the proposed method dependent on the number of visits}\label{ncanda-a}
\end{table}

\begin{figure}[t]
\centering
\subfloat[Saliency matrix derived by the proposed model]{\includegraphics[width=0.75\linewidth]{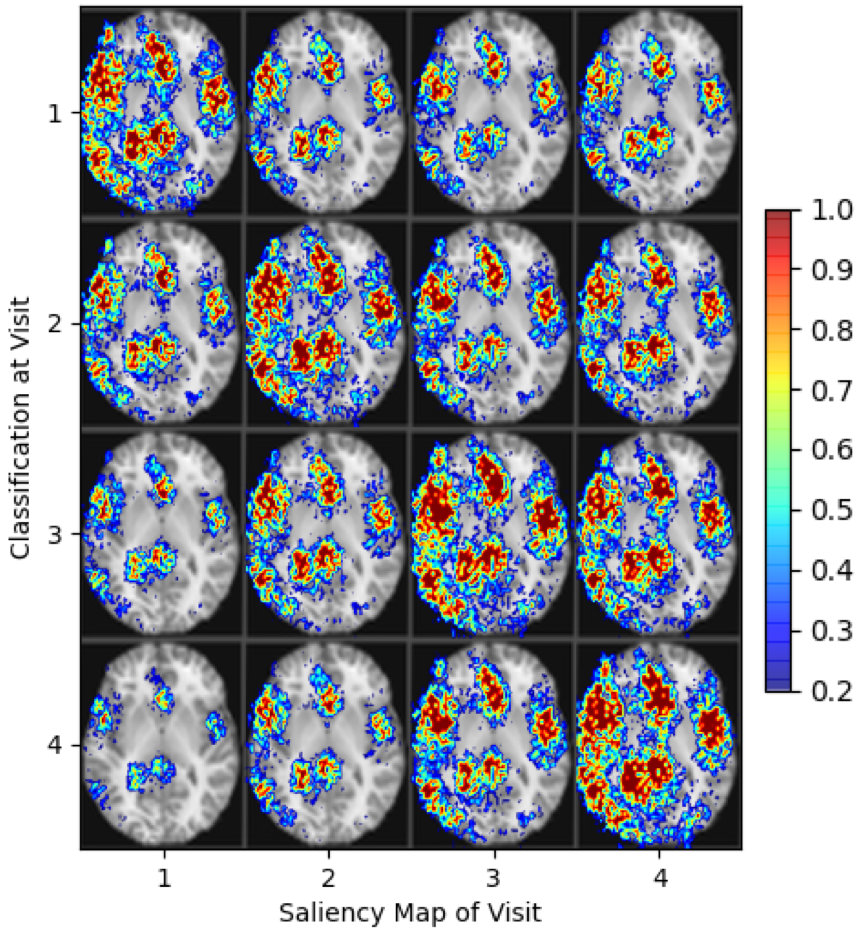}%
\label{fig:map-ncanda-matrix}}\\
\subfloat[Average saliency map of the proposed model]{\includegraphics[width=0.75\linewidth]{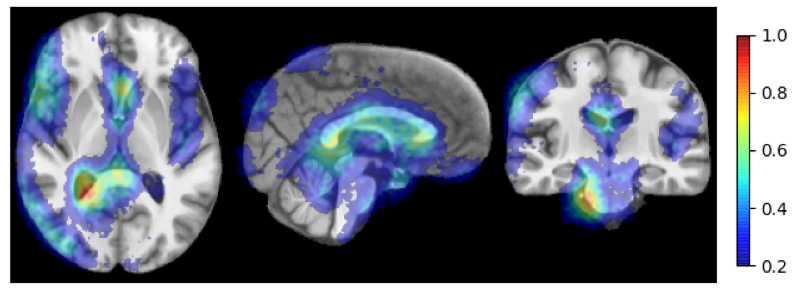}%
\label{fig:map-ncanda-mean}}
\caption{Saliency specific to youth transitioning to heavy drinking}
\label{fig:map-ncanda}
\end{figure}

Fig. \ref{fig:map-ncanda-matrix} and Fig. \ref{fig:map-ncanda-mean} both indicate that the classification was mainly driven by the inferior parietal lobule and the lateral and third ventricles. These findings comport with prior studies revealing the lateral and third ventricular are enlarged in  adult with AUD \cite{Wobrock2009,sullivan2000}. The saliency focuses more on the left hemisphere despite flipping the MRI during training. This potentially indicates that the deep model tended to discard redundant information to focus the learning on unilateral cues.

\section{Discussion and Conclusion}
We have proposed a generalized framework based on CNN and RNN to infer from longitudinal MRI the gradual deterioration of brain structure and function caused by neurological diseases and environmental influences. On the feature level, we proposed a novel longitudinal pooling layer that combined the features of a visit with a compact representation of information from proceeding visits. On the classification level, we included a consistency loss to characterize the gradual effect on brain structures across visits. The two proposed components can be easily plugged into any existing CNN-RNN architecture for improving the characterization of longitudinal trajectories. Our classification method was applied to three datasets: classifying Alzheimer's disease vs. controls, alcohol use disorder vs. controls, and adolescents who remained no-to-low drinkers vs. those who transitioned to heavy alcohol drinking. On these datasets, the proposed method achieved higher accuracy scores compared with cross-sectional and longitudinal baseline methods. The classifications for most subjects were consistent across visits so that they potentially could be used for tracking the impact of diseases and environmental factors on the brain. Important for generating those classifications was leveraging information across visits revealed by the saliency matrices. The average across saliency matrices also highlighted regions that enhanced classification and comported with findings from prior studies.


The proposed longitudinal pooling and consistency regularization provide simple means of injecting temporal dependency in features and predictions across time points, so in principle they can be used to extend any existing cross-sectional CNN architectures into longitudinal settings. As these two components are vertical improvement directions to any specific design of the CNN and RNN, we refrained from optimizing classification accuracy by extensive exploration of network architectures. Instead, our implementation of the networks relied on some of the most fundamental network components used in deep learning and still recorded reasonable prediction accuracies. Therefore, it is conceivable that the findings discussed herein are likely to generalize to more advanced network architectures \cite{oh2019classification,spasov2019parameter,ding2019deep}. Other potential strategies for further improving the classification accuracy include fusing information from multi-modality inputs, using high-resolution input images, and utilizing pre-training strategies such as convolutional autoencoders \cite{hosseini2016alzheimer}.

While the ADNI experiment focused on classifying AD patients from the controls, the proposed longitudinal components may also have the potential to improve stratification of the differential progression patterns among MCI subjects, such as separating early stages of MCI from the controls \cite{aghili2018predictive}, distinguishing progressive MCI from stable MCI \cite{cui2019rnn}, and predicting transition time from MCI to AD \cite{thung2018conversion}. Lastly, although in this paper we focused the discussion on MRI and neuroimaging applications, the proposed method does not impose any assumptions on the input data type nor the total number of time points. Therefore, the method is likely to be generalizable to other imaging modalities or clinical applications in a longitudinal setting, such as radiological abnormality detection in chest X-rays   \cite{santeramo2018longitudinal} and  lung nodule detection in CT \cite{gao2019distanced}.

\ifCLASSOPTIONcaptionsoff
  \newpage
\fi
\bibliographystyle{ieeetr}
\bibliography{mybibliography}




\end{document}